\documentclass[12pt]{article}
\usepackage{epsf,latexsym}
\usepackage{amsfonts,amssymb} 
\epsfverbosetrue
\newcommand{\de}{\hbox{\rm{d}}}

\newcommand{\pa}{\partial}

\newcommand{\ul}[1]{\underline{#1}}

\newcommand{\bb}{\begin{eqnarray}}
\newcommand{\ee}{\end{eqnarray}}
\newcommand{\eee}{\nonumber\end{eqnarray}}
\newcommand{\qq}{\quad}

\begin{document}

\font\twelve=cmbx10 at 13pt
\font\eightrm=cmr8

\thispagestyle{empty}

\begin{center}
${}$
\vspace{3cm}

{\Large\textbf{Perturbations to the Hubble diagram}} \\

\vspace{2cm}

{\large Thomas Sch\"ucker\footnote{also at Universit\'e de Provence,
France, schucker@cpt.univ-mrs.fr } (CPT\footnote{Centre de Physique
Th\'eorique\\\indent${}$\qq\qq CNRS--Luminy, Case
907\\\indent${}$\qq\qq 13288 Marseille Cedex 9,
France\\\indent${}$\qq
Unit\'e Mixte de Recherche (UMR 6207) du CNRS et des Universit\'es
Aix--Marseille 1 et 2\\
\indent${}$\qq et Sud Toulon--Var, Laboratoire affili\'e \`a la
FRUMAM (FR 2291)}), Ilhem ZouZou\footnote{also at
Universit\'e de Skikda, Algeria, izouzou1965@yahoo.fr }
(LPT\footnote{
Laboratoire de Physique Th\'eorique\\\indent${}$\qq\qq
Universit\'e Mentouri\\\indent${}$\qq\qq
25000 Constantine, Algeria
}) }

\vspace{3cm}

{\large\textbf{Abstract}}
\end{center} 

We compute the linear responses of the Hubble diagram
to small scalar perturbations in the Robertson-Walker metric and to
small peculiar velocities of emitter and receiver. We
discuss the monotonicity constraint of the Hubble diagram in the light
of these responses.

\vspace{2cm}

\vskip 1truecm

\noindent CPT-2005/P075\\
\vspace{2cm}

\section{Introduction}

In an expanding Robertson-Walker universe, the kinematics of
general relativity implies a one-to-one correspondence between the
apparent luminosity of a standard candle at rest and its red shift. This
correspondence, `the Hubble diagram', is modified by deviations from
maximal symmetry, ``anisotropies'', and by `peculiar' velocities of the
candle and its observer. We compute these modifications in linear
approximation separately for anisotropies and peculiar velocities still
in a purely kinematical context. Of course, anisotropies and peculiar
velocities are intimately related, but this relation presupposes a
gravitational dynamics, like Einstein's equation and the knowledge of
the matter content of the universe or like inflation. It also
presupposes the use of Boltzmann's equation for a gas of candles or
((super-)clusters of) galaxies.

\section{Our hypotheses}

We assume the kinematics of general relativity:
\begin{itemize}\item
  The gravitational
field is coded in a time-space metric of signature $+---$, we take the
velocity of light to be one.
\item
 Massive and massless, pointlike test particles, subject only to
gravity, follow timelike and lightlike geodesics.
\item
 Pointlike clocks, e.g. atomic clocks, are necessarily massive.
They move on timelike curves and
indicate proper time $\tau $.
\end{itemize}
We add the following cosmological hypotheses:
\begin{itemize}\item
 We assume that the
metric is Robertson-Walker with small scalar perturbations \cite{dod}, 
\bb \de \tau ^2=(1+2c)\,\de t^2-a^2(1+2b)\left[ \de\chi
^2+s^2\,\de \theta ^2+s^2 \sin ^2\theta \,\de\varphi ^2\, \right] .
\label{mt}\ee 
The scale factor $a(t)$ 
is a strictly positive function of time only, the perturbations $b$ and
$c$ are arbitrary functions on time-space, both are much smaller than
one in absolute value. The separation of the perturbations into
anisotropies and inhomogeneities makes no sense for closed universes
and by abuse we call the perturbations collectively anisotropies. We
define the function of one variable by
$s(\chi )=\sin \chi $ for the sphere, $k=1$, where $0<\chi <\pi $
describes the northern hemisphere. We put $s(\chi )=\chi $ for the
Euclidean space, $k=0$, with $0<\chi <\infty $ and
$s(\chi )=\sinh \chi $ for the pseudo-sphere, $k=-1$, with $0<\chi
<\infty $.
We take the
coordinates $\chi ,\theta ,\varphi $ dimensionless and call them
`co-moving position', while the scale factor is measured in meters.
\item
 The test particles are (superclusters of) galaxies and photons.
The former are at rest
$(t=\tau ,\ \chi ,\theta ,\varphi={\rm constant})$ plus a small
`peculiar' velocity. Note that in absence of anisotropies and peculiar
velocities, the proper time is
universal for all these timelike geodesics and is taken as time
coordinate. Under the same conditions, $\chi $ measures the
dimensionless, co-moving, geodesic distance of a position from the
origin at
$\chi =0$.
\end{itemize} 

\section{Christoffels}

We list the non-vanishing Christoffel symbols
in linear approximation in $b$ and $c$  where the underlined terms 
 refer to the symmetric case. We denote by $\pa_t$ ordinary and
partial derivative with respect to
$t$ and similarly for the other coordinates.
\vskip2mm
\begin{tabular}{cclcccr}
&&&&${\Gamma ^t}_{tt}$&=&$\pa_t c,$\\[2mm]
${\Gamma ^t}_{\chi \chi
}$&=&
\qq\qq\qq\ \ $\ul{ a\pa_t a} + 2a\pa_t a(b-c)+a^2\pa_t b,$
&&${\Gamma ^t}_{t\chi }$&=&$\pa_\chi  c,$\\[2mm]
$ {\Gamma ^t}_{\theta
\theta  }$&=& $\qq \qq \  s^2\left[ \ul{ a\pa_t a} + 2a\pa_t
a(b-c)+a^2\pa_t b\right] ,$
&& 
${\Gamma ^t}_{t\theta }$&=&$\pa_\theta  c,$
\\[2mm]
${\Gamma ^t}_{\varphi
\varphi }$&=&
$\sin^2\theta\,s^2 \left[ \ul{ a\pa_t a} + 2a\pa_t a(b-c)+a^2\pa_t
b\right] ,$&&$ {\Gamma ^t}_{t\varphi
}$&=&$\pa_\varphi  c,$
\\ [6mm]
${\Gamma ^\chi }_{tt}$&=&$a^{-2}\pa_\chi c,$&& ${\Gamma ^\chi }_{t
\chi }$&=&$\ul{\pa_t a/a}+\pa_t b, $\\[2mm]
&&&& ${\Gamma ^\chi }_{\chi \chi
}$&=&$\pa_\chi b,$
\\[2mm]
${\Gamma ^\chi }_{\theta \theta }$&=&$\qq\qq
\qq \ul{-s\pa_\chi s}-s^2\pa_\chi b,$&& 
${\Gamma ^\chi }_{\chi \theta  }$&=&$\pa _\theta  b,$\\[2mm]
${\Gamma ^\chi }_{\varphi \varphi }$&=&$\sin^2\theta
\,\left[ \ul{-s\pa_\chi s}-s^2\pa_\chi b\right] ,$&&
${\Gamma ^\chi }_{\chi \varphi   }$&=&$\pa _\varphi   b,$
\\ [6mm]
${\Gamma ^\theta  }_{tt}$&=&$a^{-2}s^{-2}\pa_\theta  c,$&& 
${\Gamma
^\theta  }_{t\theta  }$&=&$\ul{\pa_t a/a}+\pa_t b, $\\[2mm]
 ${\Gamma ^\theta  }_{\chi \chi
}$&=&$\ \ -s^{-2}\pa_\theta  b,$&&
${\Gamma ^\theta  }_{\chi \theta  }$&=&$\ul{\pa_\chi s/s}+\pa _\chi  
b,$
\\[2mm]
&& &&${\Gamma ^\theta  }_{\theta \theta }$&=&$\pa_\theta 
b,$
\\[2mm]
${\Gamma ^\theta  }_{\varphi \varphi }$&=&$
 \ul{-\sin\theta \,\cos\theta }-\sin^2\theta \,\pa_\theta  b,$&&
${\Gamma ^\theta  }_{\theta  \varphi   }$&=&$\pa _\varphi   b,$
\\ [6mm]
${\Gamma ^\varphi   }_{tt}$&=&$a^{-2}s^{-2}\sin^{-2}\theta\,
\pa_\varphi   c,$&& 
${\Gamma
^\varphi   }_{t\varphi   }$&=&$\ul{\pa_t a/a}+\pa_t b, $\\[2mm]
 ${\Gamma ^\varphi   }_{\chi \chi
}$&=&$\ \ -s^{-2}\sin^{-2}\theta\,\pa_\varphi   b,$&&
${\Gamma ^\varphi   }_{\chi \varphi   }$&=&$\ul{\pa_\chi s/s}+\pa
_\chi   b,$
\\[2mm]
${\Gamma ^\varphi   }_{\theta \theta
}$&=&$\qq\qq-\sin^{-2}\theta\,\pa_\varphi   b,$&& 
${\Gamma ^\varphi   }_{\theta  \varphi   }$&=&$\ul{\cot\theta} +\pa
_\theta    b,$\\[2mm]
&&&&${\Gamma ^\varphi   }_{\varphi \varphi }$&=&$
\pa_\varphi   b,$
\end{tabular}

\section{Anisotropies in the Hubble diagram}

The Hubble diagram is a two-dimensional parametric plot. The
parameter is the time of flight
of the photon between the emitting galaxy and receiving one, us
today. The two observables are the apparent luminosity $\ell$ and the
spectral deformation $z$. According to our model they are functions of
the unobserved time of flight, which is therefore treated as parameter
and eliminated \cite{berry}. These calculations are feasible to first
order in the perturbations of the Robertson-Walker metric.

\subsection{Trajectories of emitter and receiver}

Our first task is to compute how the trajectories of galaxies are
perturbed by the anisotropies $b$ and $c$. In the symmetric case and
without peculiar velocities these geodesics are at rest with respect
to the co-moving coordinates and of course we take the affine
parameter $p$ to be proper time $\tau $,
$t=p=\tau$,
$\chi =\chi _e$, $\theta =\theta _e$, $\varphi =\varphi _e$. 
We denote by an overdot the ordinary
derivative of the trajectory with respect to its affine parameter.
To first
order, we have
\bb \dot t =:1+\eta _t,\qq
\dot\chi =:\eta _\chi ,\qq
\dot \theta=:\eta _\theta ,\qq
\dot\varphi =:\eta _\varphi .\ee
With the Hubble rate $H:=\pa_t a/a$, the deviations $\eta _\cdot$
satisfy
\bb\dot\eta_t& +& 
\qq\qq\qq\qq\qq\qq\qq\qq\qq\pa_tc\ =\ 0,\\
\dot \eta _\chi &+&2H\eta _\chi \ +\ (\qq\qq \ a)^{-2}\pa_\chi c\ =\
0,\\
\dot \eta _\theta  &+&2H\eta _\theta  \ +\ (\qq
 \qq sa)^{-2}\pa_\theta  c\ =\ 0,\\
\dot \eta _\varphi  &+&2H\eta _\varphi  \ +\ (\sin\theta\,sa)
^{-2}\pa_\varphi  c\ =\ 0.\ee
To first order, the first equation decouples and we get
\bb \,\frac{\de t}{\de \tau }\, =1-c\left(
\tau ,\chi _e,\theta _e,\varphi _e\right) .\ee
The other three equations produce peculiar velocities,
\bb\,\frac{\de \chi }{\de\tau }\, =-\left\{\exp\!\!\left[-\int_{t_e}^\tau
2H\left( t(\tilde \tau )\right) \de\tilde \tau\right] \right\} \left\{
\int_{t_e}^{\tau}a(\hat \tau )^{-2}
\pa_\chi
\stackrel{\circ}{c}\!\!(\hat \tau )\,\exp\!\!\left[ \int_{t_e}^{\hat \tau}
2H\left( t(\tilde \tau )\right) \de\tilde \tau\right]\de\hat
\tau\right\} ,\ee 
and similarly for the perpendicular components. In this section we
will ignore peculiar velocities, the emitter is held at rest, only its
proper time is affected by the perturbations. 

The perturbed proper time of the receiver is given by a similar
formula.

\subsection{Trajectories of photons}

We solve (in first order) the geodesic
equation of  a photon  emitted from  a galaxy at time
$t_e$ and position $\chi_e ,\ \theta _e,\ \varphi _e$ and received
at time
$t_0$, today, at our position, which of course we take in the center of
the universe, $\chi =0$. Fortunately, the singularity of the metric
tensor, equation (\ref{mt}), in the center is only a coordinate
singularity. We need the link between the time of flight $t_0-t_e$ and
the geodesic distance $\chi $. 
To zeroth order in the anisotropies $b$ and $c$, the trajectory of the
photon is given by $\dot t = a_e/a$, $\dot\chi =-a_e/a^2$, $\dot \theta
=\dot\varphi =0$ with $a_e:=a(t_e)$. To first order we write
\bb \dot t =:a_e/a+\epsilon _t,\qq
\dot\chi =:-a_e/a^2+\epsilon _\chi ,\qq
\dot \theta=:\epsilon _\theta ,\qq
\dot\varphi =:\epsilon _\varphi .\ee
The geodesic equation 
becomes:
\bb && \dot\epsilon _t-\,\frac{a_e}{a}\,  H \epsilon _t-2a_eH
\epsilon _\chi  +\left( \,\frac{a_e}{a}\,\right) ^2
\pa_t(b+c)-2\,\frac{a_e^2}{a^3}\,\pa_\chi
c+2\left( \,\frac{a_e}{a}\,\right) ^2\,H(b-c)=0,\\[2mm]   &&
\dot\epsilon _\chi +2\,\frac{a_e}{a}\,H\epsilon _\chi
-2\,\frac{a_e^2}{a^3}\,\pa_tb+ \,\frac{a_e^2}{a^4}\,\pa_\chi
(b+c)=0,\\[2mm]  &&
\dot\epsilon _\theta +2\,\frac{a_e}{a}\,\left[ H-\,\frac{\pa_\chi
s}{as}\,
\right] \epsilon _\theta -\qq\qq\qq \
s^{-2}\,\frac{a_e^2}{a^4}\,\pa_\theta (b-c)=0,\\[2mm]  &&
\dot\epsilon _\varphi  +2\,\frac{a_e}{a}\,\left[ H-\,\frac{\pa_\chi
s}{as}\,
\right]
\epsilon _\varphi  -\sin^{-2}\theta
\,s^{-2}\,\frac{a_e^2}{a^4}\,\pa_\varphi  (b-c)=0.
\ee
To first order, the first two equations decouple and we get the solution
\bb \epsilon _t+a\epsilon _\chi =\,\frac{a_e}{a}\,(
\stackrel{\circ}{b} -\stackrel{\circ}{c}) ,\ee
where $\stackrel{\circ}{b}$ is the function $b$
evaluated along the zeroth order geodesic:
\bb \stackrel{\circ}{b} \!\!(p):=
b(  \stackrel{\circ}{t }\!\!(p),
\stackrel{\circ}{\chi }\!\!(p),\theta _e,\varphi _e) .\ee
The desired link between the time of flight of the photon and its
geodesic distance covered is given to first order by:
\bb \,\frac{\de \chi }{\de t}\, =-\,\frac{1}{a}\,
+\,\frac{\stackrel{\circ}{b} -\stackrel{\circ}{c}}{a}\, 
=:-\,\frac{1}{\alpha }\, .\label{chi-t}\ee
Let us rewrite this equation in terms of the emission time $t_e$,
\bb\,\frac{\de\chi }{\de t_e}\, =\,
-\,\frac{1-(b-c)(t_e,\stackrel{\circ}{\chi }\!\!(t_e),\theta _e,\varphi
_e)}{a(t_e)}\,,\qq \stackrel{\circ}{\chi }\!\!(t_e):=\int_{t_e}^{t_0}
\,\frac{\de t}{a(t)}\, \label{dchi/dte}\ee
and integrate
\bb\chi (t_e)=\stackrel{\circ}{\chi }\!\!(t_e)-
\int_{t_e}^{t_0}\,\frac{(b-c)(t,\stackrel{\circ}{\chi }\!\!(t),\theta
_e,\varphi _e)}{a(t)}\, \de t.\ee
To first order and for a fixed direction $(\theta _e,\varphi _e)$ we still
have a one-to-one correspondence between emission time and
geodesic distance. This correspondence is of course direction
dependent. 

\subsection{Spectral deformation}

Now we are ready to compute the spectral deformation of the photon
emitted at $(t_e,\chi _e, \theta _e,\varphi _e)$ with period $T_e$ 
measured by the proper time of the emitter $\tau _e$ and received at
$(t_0,0, \cdot,\cdot)$. Let us denote by $T_0$ the Doppler-shifted period
as measured by the proper time of the receiver $\tau _0$. As the period
of the photon is infinitesimal with respect to its time of flight we have
\bb \chi _e= \int_{t_e}^{t_0}\,\frac{\de t}{\alpha (t)}\, 
= \int_{t_e+T_e\,\de t/\de\tau _e}^{t_0+T_0\,\de t/\de\tau
_0}\,\frac{\de t}{\alpha (t)}\, .\label{time}\ee
Taylor expanding we obtain
\bb \,\frac{T_e\,\de t/\de\tau _e}{\alpha _e}\, =
 \,\frac{T_0\,\de t/\de\tau _0}{\alpha _0}\, \ee
and the spectral deformation,
\bb z:=\,\frac{T_0-T_e}{T_e}\, =\,\frac{a_0}{a_e}\, [
1+b_0-b_e] -1,\qq b_0:=b(t_0,0,\cdot,\cdot).\label{tof-spdef}\ee
Note that to first order the spectral deformation is independent of the
perturbation $c$. Note also that the spectral deformation now depends
on the direction via $b_e:=b(t_e,\chi _e,\theta _e,\varphi _e)$.

\subsection{Apparent luminosity}

 We
suppose known the absolute luminosity $L$ of the standard candle in
Joule per second. Our hypothesis about photons flying on geodesics
implies that the number of photons is constant. The energy $E$ of
each photon changes as its frequency $1/T$. A unit time interval
$\tilde T_e$ during which a certain number of photons are emitted is
measured by the proper time
$\tau _e$. The apparent luminosity $\ell$ is
measured in Joule per second and per square meter. Now the unit time
interval $\tilde T_0$ during which we count the received photons is
measured  by  the proper time $\tau _0$. The relation between the
unit time intervals is computed by a formula similar to equation
(\ref{time}):
\bb\,\frac{\tilde T_0}{\tilde T_e}\, =\,\frac{a_0}{a_e}\, [
1+b_0-b_e].\ee
  We also need the
(orthogonal) detector area determined by a given (infinitesimal) solid
angle
$\de\Omega $ in the direction
$\theta _e,\varphi _e$  at time $t_0$ and at
(co-moving) geodesic distance $\chi _e$. This area is measured by the
velocity of light times the proper time $\tau _0$ all squared
and is given by
\bb \de A= a_0^2\,s^2(\chi _e)\,[1+2b_0] \,\de
\Omega  .\ee
Note that to first order in $b$ and $c$ we may still speak about photons
propagating in a given solid angle. Finally the apparent luminosity
is:
\bb\ell=\,\frac{L}{4\pi}\,\frac{\de \Omega }{\de A}\, 
\frac{E_0}{E_e}\,\frac{\tilde T_e}{\tilde T_0}\,=
\,\frac{L}{4\pi\, a_0^2\,s^2(\chi _e) }\,
\left( \frac{ a_e}{ a_0}\right) ^2\, [1-4b_0+2b_e].
\ee
Note that as for the spectral deformation, the apparent luminosity to
first order does not depend on the perturbation $c$ but does depend on
the direction.

\subsection{Eliminating the time of flight}

Our last task is the elimination of the unobserved parameter, the time
of flight.
To this end we differentiate the relation between time of flight and
spectral deformation, equation (\ref{tof-spdef}),
\bb z(t_e)+1=\,\frac{a_0}{a_e}\,\left[ 
1+b_0-b(t_e,\stackrel{\circ}{\chi }\!\!(t_e),\theta _e,\varphi
_e)\right]  ,\ee
with respect to $t_e$:
\bb\,\frac{\de z}{\de t_e}\, =&-&\frac{a_0}{a(t_e)^2}\, \dot a(t_e)
\,\left[ 
1+b_0-b(t_e,\stackrel{\circ}{\chi }\!\!(t_e),\theta _e,\varphi
_e)\right]\cr \cr 
 &+&\frac{a_0}{a(t_e)}\,\left[ 
-\pa _t b(t_e,\stackrel{\circ}{\chi }\!\!(t_e),\theta _e,\varphi
_e)+\pa _\chi  b(t_e,\stackrel{\circ}{\chi }\!\!(t_e),\theta _e,\varphi
_e)/a(t_e)\right] \,.\ee
From this  and equation (\ref{dchi/dte}) we get
\bb\,\frac{\de \chi }{\de z}& =&\frac{\de \chi }{\de t_e}\,
/\,\frac{\de z}{\de t_e}\,\cr \cr 
& =&\,\frac{1}{a_0H(z)}\, \left[ 
1-b_0+c(t_e(z),\stackrel{\circ}{\chi }\!\!(z),\theta _e,\varphi
_e)\right.\cr  &&\left.\qq\qq\qq\qq\qq
-H(z)^{-1}(\pa_t-a(z)^{-1}\pa_\chi
)b(t_e(z),\stackrel{\circ}{\chi }\!\!(z),\theta _e,\varphi _e)\right] 
\ee
with
\bb \stackrel{\circ}{\chi }\!\!(z)=\,\frac{1}{a_0}\,
\int_0^z\,\frac{\de\tilde z}{H(\tilde z)}\,  \ee
and integrating
\bb \chi (z)=\stackrel{\circ}{\chi }\!\!(z) +\delta ,\ee
\bb\delta (z) :=-b_0\stackrel{\circ}{\chi }\!\!(z)
&+&\int_0^z\,\frac{c(t_e(\tilde z),\stackrel{\circ}{\chi }\!\!(\tilde
z),\theta _e,\varphi _e)}{a_0H(\tilde z)}\, \de\tilde z\cr 
&-&\int_0^z\,\frac{(\pa_t-a(\tilde z)^{-1}\pa_\chi)b(t_e(\tilde z),
\stackrel{\circ}{\chi }\!\!(\tilde z),\theta _e,\varphi _e)}{a_0H(\tilde
z)^2}\, \de\tilde z.\ee

\subsection{Hubble diagram}

Finally the Hubble diagram is to first order in the scalar perturbations
$b$ and $c$:
\bb \ell (z)=\,\frac{L}{4\pi  a_0^2(z+1)^2s^2(\stackrel{\circ}{\chi
}\!\!(z))}\, \left[ 1-2b_0-2\,\frac{s'}{s} (\stackrel{\circ}{\chi
}\!\!(z))\, \delta (z) \right]. \ee
Our unit of time is chosen today and here on earth. Therefore we set
$c_0=0$. Likewise our unit of length or more precisely the numerical
value of the speed of light is chosen here and now and we set $b_0=0$.
This shows that the apparently strongest $z$-dependence of the
linear correction to the Hubble   stemming from the term $-b_0\!\!
\stackrel{\circ}{\chi}$ is a coordinate artifact. The remaining terms
are weighted averages of $c$ and a derivative of $b$ along the zeroth
order path of the photon between the standard candle and us today. In
the absence of particular conspiracies in the perturbations, the
$z$-dependence of these terms is weak:
\bb \delta (z)=z\left[ 
\,\frac{c(t_e( z_{\rm int}),\stackrel{\circ}{\chi }\!\!(
z_{\rm int}),\theta _e,\varphi _e)}{a_0H( z_{\rm int})}
-\,\frac{(\pa_t-a( z_{\rm int})^{-1}\pa_\chi)b(t_e( z_{\rm int}),
\stackrel{\circ}{\chi }\!\!( z_{\rm int}),\theta _e,\varphi
_e)}{a_0H( z_{\rm int})^2}\, \right] \ee
for some intermediate value $z_{\rm int}\in [0,z]$. Indeed, a recent fit
to the Hubble diagram \cite{ff} up to $z=1.8$  gives
$H(z)=H_0(z+1)^{0.69}$. 

We conclude that the scalar perturbations produce a Hubble diagram
which is a band in the $z\ell$ plane with more or less constant
relative vertical width. This relative width is of the same order of
magnitude as the perturbations.

\section{Peculiar velocities in the Hubble diagram}

In this section our metric is Robertson-Walker without perturbations,
$b=c=0$. However we admit peculiar velocities of emitter and receiver
with respect to the co-moving coordinates $\chi ,\ \theta ,\ \varphi $,
or, put more physically, with respect to the cosmic microwave
background. We compute the changes in the Hubble diagram to first
order in the peculiar velocities divided by the speed of light, which
we have set to one.

\subsection{Trajectories}

We take the line of sight in the direction $\theta =\pi /2,\ \varphi
=0,$ and decompose the peculiar velocities into parallel and
perpendicular components with respect to this direction: 
$\vec v_e=\vec v_{e\parallel}+\vec v_{e\perp}$,  
$\vec v_0=\vec v_{0\parallel}+\vec v_{0\perp}$. Then we get the
initial conditions of the emitter at $t=t_e$
\bb \dot t = \,\frac{\de t}{\de \tau _e}\, =\sqrt{1+v_e^2},\qq
\dot \chi =\,\frac{v_{e\parallel}}{a_e}\, \qq
\dot\theta =0,\qq
\dot \varphi =\,\frac{v_{e\perp}}{a_es_e}\, ,\ee
 the initial conditions of the receiver at $t=t_0$
\bb \dot t = \,\frac{\de t}{\de \tau _0}\, =\sqrt{1+v_0^2},\qq
\dot \chi =\,\frac{v_{0\parallel}}{a_0}\, \qq
\dot\theta =0,\qq
\dot \varphi =\,\frac{v_{0\perp}}{a_0s_0}\, ,\ee
and the initial conditions of the go-between at $t=t_e$
\bb \dot t = \,\frac{\de t}{\de p}\, =1,\qq
\dot \chi =-\,\frac{1}{a_e}\, \qq
\dot\theta =0,\qq
\dot \varphi =0 .\ee
The connection between geodesic distance covered by the photon and
its time of flight is
\bb \chi _e=\int_{t_e}^{t_0}\,\frac{\de t}{a(t)}\, .\ee

\subsection{Spectral deformation}

To compute the spectral deformation we have a second photon emitted
a period $T_e$ later with respect to the proper time $\tau _e$ of the
emitter. Therefore this photon will be emitted at
$t=t_e+\sqrt{1+v_e^2}T_e$ and at position
$\chi =\chi _e+v_{e\parallel}T_e/a_e$.
It will be received at $t=t_0+\sqrt{1+v_0^2}T_0$ and at position
$\chi =v_{0\parallel}T_0/a_0$. We therefore have
\bb \chi _e+\,\frac{v_{e\parallel}T_e}{a_e}\,
-\,\frac{v_{0\parallel}T_0}{a_0}\, 
=\int_{t_e+\sqrt{1+v_e^2}T_e}^{t_0+\sqrt{1+v_0^2}T_0}\,
\frac{\de t}{a(t)}\, .\ee
Taylor expanding as before yields
\bb \,\frac{T_e}{T_0}\, =\,\frac{a_e}{a_0}\, 
\frac{\sqrt{1+v_0^2}+v_{0\parallel}}{\sqrt{1+v_e^2}+v_{e\parallel}}\,.
\ee 
To first order, the spectral deformation,
\bb z=\,\frac{a_0}{a_e}\, (1-(v_{0\parallel}-v_{e\parallel})) -1,\ee
depends only on the difference of the parallel components of the
peculiar velocities.

\subsection{Apparent luminosity}

We suppose that our candle emits its absolute luminosity $L$
isotropicly in its rest frame. When moving with velocity $\vec
v_e-\vec v_0$, its emission profile  with respect to the receiver is
\bb\,\frac{\de N}{\de \Omega }\, =\,\frac{L}{4\pi }\,
\,\frac{1-|\vec v_e-\vec v_0|^2}{(1-|\vec v_e-\vec v_0|\cos\varphi
)^2}\, \sim\,\frac{L}{4\pi }\,
\,\frac{1}{(1-(v_{0\parallel}-v_{e\parallel})
)^2}\, ,\label{profile}\ee where $\varphi $ is the angle between the
line of sight and
$\vec v_e-\vec v_0$. Our convention of orientation is such
that $\varphi =0$ and
$v_{0\parallel}-v_{e\parallel}:=|\vec v_e-\vec
v_0|\cos\varphi$ positive when the emitter moves towards the
receiver in which case the forward emission is enhanced. The
deformed emission profile, the first of equations (\ref{profile}), is a
special relativistic formula and contains a difference of velocities at
different points. Its first order approximation, the second part of
(\ref{profile}), only contains projections of velocities onto a geodesics
and makes sense also in general relativity.

Neglecting Lorentz contractions, which are quadratic in velocity, the
detector area seen by the first photon is 
$\de A=a^2_0s^2(\chi _e)\de\Omega $, 
while the second photon sees
\bb\de A=a^2_0\,s^2\!\!\left(\, \chi
_e+\,\frac{v_{e\parallel}T_e}{a_e}\,
-\,\frac{v_{0\parallel}T_0}{a_0}\, \right)\,\de\Omega\ 
\sim\  a^2_0s^2(\chi _e)\left[ 
1-2(v_{0\parallel}-v_{e\parallel})\,\frac{T_0s'(\chi _e)}{a_0s(\chi
_e)}\,\right]\de\Omega  . \ee
The term $T_0/(a_0s_e)$ is an atomic period divided by the time of
flight and can safely be dropped.
 Note also that we do not have to worry about  the
angle between the detection area and the line of sight which for a
moving observer optimizing her  efficiency deviates from $90^\circ$
by an amount quadratic in her velocity. Therefore to first order the
apparent luminosity is:
\bb\ell=\,\frac{\de N}{\de\Omega }\,\frac{\de \Omega }{\de A}\, 
\frac{E_0}{E_e}\,\frac{\tilde T_e}{\tilde T_0}\,=
\,\frac{L}{4\pi\, a_0^2\,s^2(\chi _e) }\,
\left( \frac{ a_e}{ a_0}\right) ^2[1+4(v_{0\parallel} -
 v_{e\parallel})].
\ee

\subsection{Hubble diagram}

We eliminate the time of flight as in the preceding section and get
the Hubble diagram with its linear perturbations coming from the
peculiar velocities of emitter and receiver:
\bb \ell (z)=\,\frac{L}{4\pi  a_0^2(z+1)^2s^2(\stackrel{\circ}{\chi
}\!\!(z))}\, \left[ 1+2\sigma (
\stackrel{\circ}{\chi }\!\!(z)) (v_{0\parallel} -
 v_{e\parallel})\right].
\ee
The function $\sigma (\chi ):=1-\chi s'(\chi)/s(\chi )$ vanishes
identically for flat universes, $k=0$. For curved universes,
$k=\pm 1$,
$\sigma (\chi )$ is small,
\bb \sigma (\chi ) = \,\frac{k}{3}\, \chi ^2 +
\,\frac{1}{45}\, \chi ^4+\,\frac{2k}{945}\, \chi ^6\pm ...\ee

We conclude that peculiar velocities do not perturb the Hubble
diagram to first order if the universe is flat. For curved universes the
linear perturbation is small for small redshift and grows with $z$.

\section{Conclusions}

Under the very general kinematical hypotheses outlined at the
beginning, the unperturbed Hubble diagram has a monotonicity
property  \cite{andre}. Today, this property is respected by
supernova data. That might change in a foreseeable future and we
already look for excuses. The first two that come to mind are
fluctuations in the absolute luminosity $L$ and absorption by dirt
along the line of sight. We find it hard to believe that these two effects
show a $z$ dependence that will mimic a non-monotonicity in the
Hubble diagram. We rather expect that they will produce a band in the
$z\ell$ plane (single side band for dirt). Two other excuses come to
mind next, anisotropies and peculiar velocities. After the above
calculations, we find it hard to believe, that anisotropies and peculiar
velocities can account for violations of the monotonicity constraint in
the Hubble diagram.

\end{document}